\documentclass[aps,prl,twocolumn,showpacs]{revtex4-1}
\usepackage{epsfig}
\usepackage{graphicx}
\usepackage{bm}
\usepackage{amssymb}
\usepackage{amsmath}
\usepackage{amsthm}
\begin{document}
\title{Universality classes and crossover scaling of Barkhausen noise in thin films}
\author{Lasse Laurson$^{1}$, Gianfranco Durin$^{2,3}$, and Stefano Zapperi$^{2,4}$}
\affiliation{$^1$COMP Centre of Excellence, Department of Applied Physics, Aalto University, 
P.O. Box 14100, FIN-00076, Aalto, Espoo, Finland}
\affiliation{$^2$ ISI Foundation, Via Alassio 11/c 10126 Torino, Italy}
\affiliation{$^3$Istituto Nazionale di Ricerca Metrologica, strada delle Cacce 91, 10135 
Torino, Italy}
\affiliation{$^4$CNR - Consiglio Nazionale delle Ricerche, IENI, Via R. Cozzi 53, 20125, 
Milano, Italy}
\begin{abstract}
We study the dynamics of head-to-head domain walls separating in-plane domains in a 
disordered ferromagnetic thin film. The competition between the domain wall surface
tension and dipolar interactions induces a crossover between a rough domain wall phase
at short length-scales and a large-scale phase where the walls display a zigzag morphology. 
The two phases are characterized by different critical exponents for Barkhausen avalanche dynamics
that are in quantitative agreement with experimental measurements on MnAs thin films.
\end{abstract}
\maketitle

When subject to an external magnetic field, a ferromagnetic material shows a
sequence of discrete and intermittent jumps of the magnetic domain walls (DW's),
known as the Barkhausen effect \cite{DUR-06}, a paradigmatic example of
crackling noise in materials \cite{SET-01}. The statistical properties of the Barkhausen noise 
are usually studied by measuring the size distribution $P(s)$ of such jumps, or avalanches, 
which typically follows a power law $P(s) \sim s^{-\tau}$, with the exponent 
$\tau$ characterizing the universality class of the avalanche dynamics. 
In three dimensional bulk ferromagnetic materials, the scaling behavior 
of the Barkhausen effect is understood theoretically in
terms of the depinning transition of domain walls \cite{ZAP-98} with two 
distinct universality classes for amorphous and polycrystalline materials \cite{DUR-00}.
A similar clear-cut classification does not exist in lower dimensions, despite Barkhausen avalanches 
having been studied experimentally for decades in several ferromagnetic thin films with in-plane 
\cite{WIE-77,WIE-78,PUP-00,KIM-03,SAN-06,RYU-07}
or out-of-plane anisotropy \cite{SCH-04,LIE-05}. This issue is particularly important because these 
low-dimensional magnetic structures have become increasingly relevant for
various technological applications \cite{PAR-08,HAY-08}.

An important step towards understanding the different universality classes in thin
films was achieved by the magneto-optical experiments of Ryu {\it et al.} 
\cite{RYU-07}, who observed a crossover between two different avalanche size 
exponents $\tau$ as temperature $T$ was varied close to but below the Curie 
temperature $T_c$ of a 50 nm MnAs film. This crossover was accompanied 
by changes in DW morphology, such that the DW structure evolves from rough for high $T$ 
to DW's with a pronounced tendency to form zigzag or sawtooth -like patterns for lower 
$T$. It was argued that by varying $T$ close to $T_c$, one can tune the value of the 
squared saturation magnetization $M_s^2$, and thus the strength of the long-range 
dipolar interactions between different DW segments. The zigzag pattern is expected to 
arise as a result of a competition between the domain wall energy and the 
dipolar interactions, with the former favoring a flat horizontal DW, while the latter 
would prefer a vertically spread DW to reduce the magnetic charge density 
\cite{MUG-10,CER-06,CER-07}. 

In this Letter, we provide a theoretical explanation for the experimentally observed  
universality classes and the crossover between them.  Due to the essentially $2d$ thin film geometry 
considered here (the film thickness $\Delta_z$ is much smaller than the DW length), we 
model the DW as a flexible line $\Sigma$ with surface tension $\gamma_w$ due to DW 
energy. The line moves within the $xy$ plane, and has an average orientation along 
the $x$ axis. It is taken to separate two magnetic domains with magnetization along 
$\pm \hat{{\bf y}}$, respectively. Thus, a head-to-head DW is characterized by a 
magnetic charge density $\sigma({\bf r}) = 2M_s \cos \theta({\bf r})$ along the DW, 
with $\theta ({\bf r})$ the angle between the local DW normal $\hat{\bf n}$ and the 
$\hat{{\bf y}}$ direction. These magnetic charges then lead to a magnetostatic field 
${\bf H}_m({\bf r}) = \int_{\Sigma'} \sigma({\bf r'})({\bf r}-{\bf r'})/|{\bf r}-{\bf r'}|^3 ds'$, 
the $y$ component of which is acting on the DW segments, along with an applied field 
${\bf H}_a = H_a \hat{\bf{y}}$. In addition, the DW segments interact with quenched 
disorder, described by a random pressure field ${\bf \eta}({\bf r})$ due to short range 
interactions with random pinning centers. Thus, the total normal pressure difference
$\Delta p$ acting across the DW at point ${\bf r}$ reads 
\begin{eqnarray}
\Delta p({\bf r}) & = & \gamma_w/R({\bf r}) + 2M_s\mu_0 H_a + 
{\bf \eta}({\bf r})\cdot \hat{\bf n} + \\ \nonumber     
& & + 4 \mu_0 M_s^2 \int_{\Sigma ({\bf r}')} 
\frac{(y-y')\cos \theta'}{[(x-x')^2+(y-y')^2]^{3/2}}ds',
\end{eqnarray}
where $R({\bf r})$ is the local radius of curvature. To simulate such a system, 
we discretize the DW along the $x$ direction, by using the film thickness $\Delta_z$
as the lattice constant, and describe the DW by a single-valued function $y=h(x,t)$, 
with $x=i=1,2,\dots, L$. The local DW velocity is assumed to be proportional to the 
local pressure acting on the DW, such that the equation of motion for the DW line 
segment $i$ along the $y$ direction is given by
\begin{eqnarray}
\label{eq:eom}
\Gamma \frac{\partial h_i}{\partial t} & = & \frac{1}{\cos \theta_i} 
\bigg[ \gamma_w \frac{\partial^2 h_i}{\partial x^2} + 
2M_s\mu_0 H_a + \eta(i,h_i) + \\ \nonumber
& & + 4\mu_0 M_s^2\Delta_z^2 
\sum_{j \neq i} \frac{h_i-h_j}{[\Delta_z^2(i-j)^2+(h_i-h_j)^2]^{3/2}} \bigg],
\end{eqnarray}
where we have approximated the curvature term by a discretized Laplacian, $\theta_i$ is the 
angle between the normal of the $i$th segment and the $y$ direction, and $\Gamma$ is a
damping constant. The factor $1/\cos \theta_i$ multiplying the right hand side of 
Eq. (\ref{eq:eom}) transforms normal motion into motion along the $y$ direction. 
The quenched random force has correlations 
$\langle \eta(i,h_i) \eta(j,h_j)\rangle = \sigma^2 \delta (i-j) \delta (h_i - h_j)$.
We further write Eq. (\ref{eq:eom}) in non-dimensional units, by  
measuring lengths in units of $\Delta_z$ and times in units of 
$\Gamma \Delta_z/(\mu_0 M_s^2)$. The resulting dimensionless equation of motion reads
\begin{eqnarray} 
\label{eq:eomnd}
\frac{\partial h_i}{\partial t} & = & \frac{1}{\cos \theta_i} 
\bigg[\lambda\frac{\partial^2 {h}_i}{\partial {x}^2} + 
 {F}_{ext} +  {\eta}(i, {h}_i) + \\ \nonumber
& & + 4\sum_{j \neq i} \frac{ {h}_i- {h}_j}{[(i-j)^2+( {h}_i- {h}_j)^2]^{3/2}} \bigg],
\end{eqnarray}
where the dimensionless driving force is 
${F}_{ext} = 2H_a/M_s$ and $\lambda \equiv l_D/\Delta_z$ is the ratio between the 
``domain formation'' length \cite{BER-98} $l_D=\gamma_w/(\mu_0 M_s^2)$ and the film 
thickness. In dimensionless units, the quenched random force has correlations 
$\langle  {\eta}(i, {h}_i)  {\eta}(j, {h}_j)\rangle =
 {\sigma}^2 \delta (i-j) \delta ( {h}_i -  {h}_j)$, with
$ {\sigma} = \sigma/(\mu_0 M_s^2 \Delta_z)$.
Periodic boundary conditions are implemented by using the nearest image approximation
to compute the non-local dipolar forces.

\begin{figure}[t!]
\includegraphics[width=7cm,clip]
{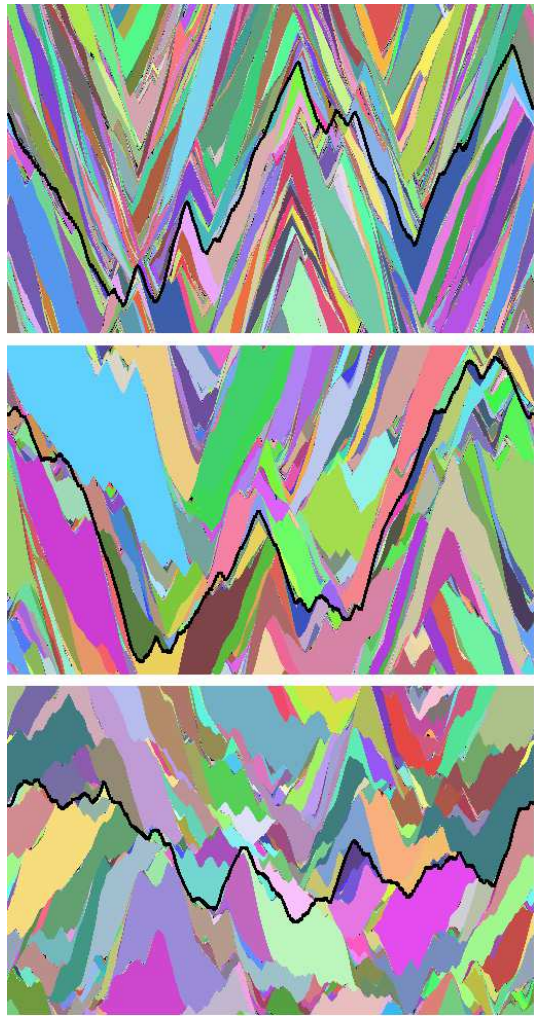}
\caption{(color online) The spatial structure of Barkhausen
avalanches for $  \lambda = 1$ (top), $  \lambda = 2$ (middle) 
and $  \lambda = 4$ (bottom).
The domain wall is moving from top to bottom, and the area swept 
over by each avalanche has been colored with a random color.
An example of the DW structure is given by a black line in each 
case.}
\label{fig:color_avs}
\end{figure}

\begin{figure}[t!]
\includegraphics[width=8cm,clip]
{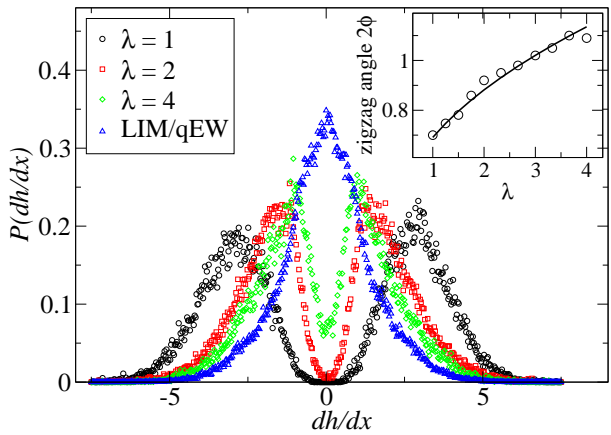}
\caption{(color online) Distributions of the
local slopes $dh/dx$ (see text for definition) for various $  \lambda$. 
The inset shows the corresponding zigzag angle $2\phi$ as a function of
$  \lambda$. The solid line is a fit to of Eq. (\ref{eq:angle}),
corresponding to $ {l}=3.9$.}
\label{fig:slopes}
\end{figure}

To mimic the experiments of Ref. \cite{RYU-07}, we simulate the system by integrating 
Eq. (\ref{eq:eomnd}) numerically, fixing the external force $ {F}_{ext}$ to a 
constant value below the critical depinning force $ {F}_c$, and monitor the 
dynamics of the DW. Whenever the average DW velocity $V(t)=1/L \sum_i \partial 
 {h}_i/\partial  {t}$ falls below a low threshold value $V_{th}$, a randomly
selected DW segment is given a ``kick'', such that an additional local force acting 
on the DW segment is first increased linearly from zero until $V>V_{th}$, and then 
decreased continuously back to zero. This can then trigger an avalanche, which lasts 
until the average velocity of the front again falls below $V_{th}$, and the process 
is repeated. The area (measured in units of $\Delta_z^2$) over which the DW moves 
between two such triggering events (which mimic the effect of thermal activation) is 
taken to be the avalanche size $ {s}$. Fig. \ref{fig:color_avs} shows the 
spatial structure of the avalanches for different $  \lambda$-values. For small
$  \lambda$, the DW's exhibits a clear zigzag morphology (with avalanches tilted
accordingly), and roughen due to disorder as $  \lambda$ is increased.

We further characterize the zigzag morphology by considering the distributions of the
local slopes $\partial  {h}/\partial  {x}$ of the DW, see Fig. \ref{fig:slopes}. 
For finite $  \lambda$, the distributions are bimodal, reflecting the fact that the dipolar 
interactions render the flat DW unstable. For the sake of comparison, we show also the 
slope distribution for the Linear Interface Model (LIM)/quenched Edwards-Wilkinson 
(qEW) equation (i.e. Eq. (\ref{eq:eomnd}) without the non-local term, corresponding to the 
limit $  \lambda \rightarrow \infty$), displaying a single peak at 
$\partial  {h}/\partial {x}=0$. 
The inset of Fig. \ref{fig:slopes} shows the zigzag angle $2\phi$, defined as
$2\phi = 2\tan^{-1}(1/\langle |\partial  {h}/\partial  {x}| \rangle)$. For 
small $  \lambda \sim 1/M_s^2$, $2\phi$ is linear in $  \lambda$, similarly to 
experimental results \cite{RYU-06}, while for very large $  \lambda$ the DW becomes 
rough, and the concept of the zigzag angle is ill-defined. An approximate analytical 
estimate of the $  \lambda$-dependence of $2\phi$ can be obtained by requiring balance 
between forces due to line tension and dipolar interactions. The former can be estimated 
as $  \lambda \partial^2  {h}/\partial  {x}^2 =   \lambda 2m/ {l}$, 
where $m=\langle |\partial  {h}/\partial {x}| \rangle$ 
is the magnitude of the zigzag slope, and $ {l}$ is the length of the 
``transition region'' at the tip of the zigzag where a constant curvature $2m/ {l}$ 
is assumed. These have to be balanced by forces due to dipolar interactions, which we 
write in terms of the slope $m$ as 
\begin{equation}
4 \sum_{j \neq i}\frac{m|i-j|}{|i-j|^3(1+m^2)^{3/2}} = 4\frac{m}{(1+m^2)^{3/2}}2\zeta (2),
\end{equation} 
where $\zeta (2) = \pi^2/6$. Thus, from the force balance condition, one obtains for 
the slope $m = \sqrt{(2 \pi^2 l/3   \lambda)^{2/3}-1}$, corresponding to the zigzag
angle 
\begin{equation}
2\phi = 2\tan^{-1}(m^{-1}) =  2\tan^{-1}[(2 \pi^2 l/3   \lambda)^{2/3}-1]^{-1/2}.
\label{eq:angle}
\end{equation}
A good fit to the data with Eq. (\ref{eq:angle}) can be obtained by using $ {l}$ 
as a fitting parameter, resulting in $ {l} \approx 3.9$, see the inset of Fig. 
\ref{fig:slopes}.

\begin{figure}[t!]
\includegraphics[width=8cm,clip]
{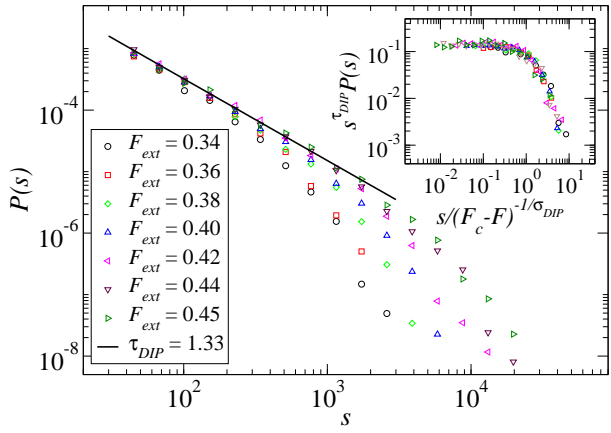}
\caption{(color online) The distribution of avalanche sizes 
$ {s}$ with $  \lambda=1$, corresponding to the limit dominated
by dipolar interactions, for various $F_{ext} \leq F_c$. The inset 
shows a collapse with exponents $\tau_{DIP} = 1.33$ and $1/\sigma_{DIP} = 3.5$.}
\label{fig:avdistLD1}
\end{figure}

For small $  \lambda$, the statistical properties of the Barkhausen avalanches 
are expected to reflect the dominant nature of the dipolar interactions.
Fig. \ref{fig:avdistLD1} shows the avalanche size distributions $P( {s})$ for 
$  \lambda=1$ and various $ {F}_{ext} <  {F}_c$. The distributions
are found to obey
\begin{equation} 
\label{eq:dipolar_scaling}
P( {s}) =  {s}^{-\tau_{DIP}}\mathcal{F}_{DIP}\left[\frac{ {s}}
{( {F}_c -  {F}_{ext})^{-1/\sigma_{DIP}}}\right],
\end{equation}
where $\mathcal{F}_{DIP}(x)$ is a scaling function, $\tau_{DIP} \simeq 1.33$ and 
$1/\sigma_{DIP} \simeq 3.5$. The value of $\tau_{DIP}$ characterizes the ``zigzag''
universality class dominated by dipolar interactions, and is close to
that found for certain other systems with long-range
anisotropic interaction kernels, such as models of amorphous
plasticity \cite{TAL-11}. For larger $\lambda$, 
while large enough avalanches are still dominated by the dipolar 
interactions, small avalanches start to be governed by the surface 
tension, and the power law part of Eq. (\ref{eq:dipolar_scaling}) has to be 
replaced by a crossover scaling form including two different power laws with 
the corresponding $\tau$-exponents,
\begin{equation}
P( {s}) =  {s}^{-\tau_{LIM}}
[1+( {s}/ {s}_{\chi})^{(\tau_{DIP}-\tau_{LIM})k}]^{-1/k}
\mathcal{F}( {s}/ {s}_0),
\label{eq:xoverfit}
\end{equation}
where $ {s}_{\chi}$ is a crossover avalanche size separating the two regimes, 
$k$ controls the sharpness of the crossover and ${s}_0$ is the cut-off avalanche 
size. The short length scale exponent is expected to be that of the LIM/qEW, 
$\tau_{LIM} \simeq 1.11$ \cite{ROS-09} and $1/\sigma_{LIM}=3.0$ \cite{ZAP-98}. 
 
To estimate the crossover scale $ {L}_{\chi}$ (and the corresponding crossover 
avalanche size $ {s}_{\chi}$) above which the dipolar forces will dominate the 
line tension, we consider the continuum version of Eq. (\ref{eq:eomnd}) for small 
deformation of the DW without disorder and external force, 
\begin{equation}
\label{eq:eomsd}
\frac{\partial  {h}}{\partial  {t}} =   \lambda 
\frac{\partial^2  {h}}{\partial  {x}^2} +
4 \int \frac{ {h}( {x}) -  {h}( {x}')}{| {x}- {x}'|^3} d {x}',
\end{equation}
and examine the stability of a flat DW. By writing the two interaction terms in Eq. 
(\ref{eq:eomsd}) in terms of their Fourier transforms, 
$  \lambda \frac{\partial^2  {h}_i}{\partial  {x}^2} 
= \int d {q}  {h}_{ {q}} e^{i 2\pi  {q}  {x}}(-4\pi^2  \lambda  {q}^2)$ 
and $4\int d {x} \frac{ {h}( {x})- {h}( {x}')}{| {x}- {x}'|^3}
= 4\int d {q} e^{i 2\pi  {q} {x}} {h}_{ {q}} \int d {x}' 
\frac{1-e^{i 2\pi  {q}( {x}'- {x})}}{| {x}- {x}'|^3}$, one
arrives at a stability condition for the mode $ {q}$, 
$-4\pi^2  \lambda  {q}^2 + I( {q})<0$, where $I( {q}) \equiv 
4\int d {r} \frac{1-e^{i 2\pi  {q} {r}}}{| {r}|^3}$.
We expand $I( {q})$ for small $ {q}$, such that $I( {q}) \simeq 
4\int_{1}^{1/ {q}}d {r} ( {q} {r})^2/| {r}|³
= -8\pi^2 {q}^2 \log( {q})$. Thus, the stability condition becomes 
$2\log ( {q})+  \lambda>0$, which leads to a crossover length 
\begin{equation}
\label{eq:xover}
 {L}_{\chi} = e^{  \lambda/2}.
\end{equation}
The crossover avalanche size is expected to scale as $ {s}_{\chi} \sim 
 {L}_{\chi}^{1+\zeta_{\chi}}$, where $\zeta_{\chi}$ is the roughness exponent of 
the avalanches at the crossover scale. Thus, also the crossover avalanche size is 
an exponential in $  \lambda$, 
\begin{equation}
\label{eq:s_xover}
 {s}_{\chi}=e^{(1+\zeta_{\chi})  \lambda/2}.
\end{equation}
Notice that this form is different from the one employed in Ref. \cite{RYU-07}.

\begin{figure}[t!]
\includegraphics[width=8cm,clip]
{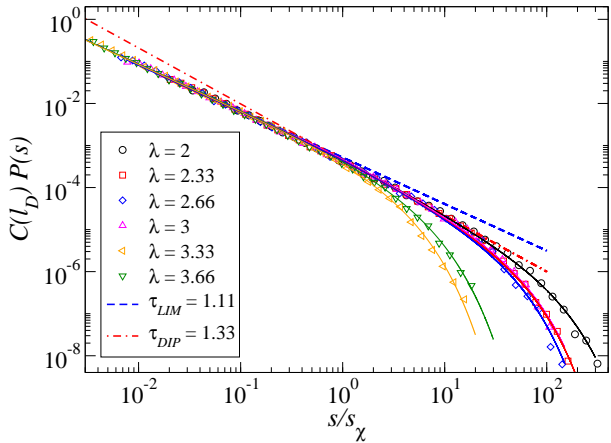}
\caption{(color online) The rescaled avalanche size distributions 
for various $  \lambda$, with $F$ slightly below $F_c$ in each case, 
showing the crossover between the two scaling exponents, $\tau_{LIM}=1.11$ 
(dashed blue line) and $\tau_{DIP}=1.33$ (dash-dotted red line). The solid
lines are fits of Eq. (\ref{eq:xoverfit} to the data. The dependence 
of the crossover avalanche size $s_{\chi}$ on $  \lambda$ resulting 
from the fits is reported in Fig. \ref{fig:crossover2} (a).) 
}
\label{fig:crossover}
\end{figure}

\begin{figure}[t!]
\includegraphics[width=8cm,clip]
{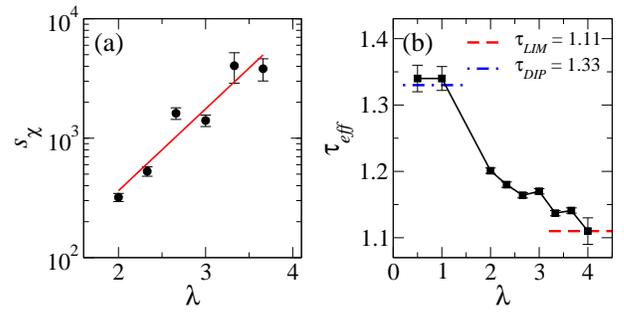}
\caption{(color online) (a) shows the exponential dependence 
of the crossover avalanche size $ {s}_{\chi}$ on $  \lambda$. The 
solid line corresponds to an exponential fit $ {s}_{\chi} = A e^{B  \lambda}$, 
with $A=15.5$ and $B=1.577$. (b) shows the effective exponent $\tau_{eff}$
as obtained by fitting Eq. (\ref{eq:single_PL}) to the data.}
\label{fig:crossover2}
\end{figure}

To test this argument, 
we simulate the model for various $\lambda \geq 1$, and 
estimate $ {s}_{\chi}(  \lambda)$ by fitting Eq. (\ref{eq:xoverfit}) to
the data. We found that $\mathcal{F}(x)=\exp(-x)$, $k=10$ (corresponding to
a sharp crossover), and $\tau_{LIM}=1.11$ and $\tau_{DIP}=1.33$ produce a very 
good fit, see Fig. \ref{fig:crossover}. Fig. \ref{fig:crossover2} (a) shows the 
resulting $ {s}_{\chi}(  \lambda)$-data, which can be well fitted by an 
exponential, thus confirming the functional form in Eq. (\ref{eq:s_xover}).
 Fig. \ref{fig:crossover} shows the avalanche size distributions 
for different $  \lambda$, with $ {s}$ rescaled with the corresponding 
$ {s}_{\chi}(  \lambda)$ and $P(s)$ by the factors $C(  \lambda)$,
chosen to make the different distributions overlap. This procedure reveals
a clear crossover scaling, with the exponents $\tau_{LIM} \simeq 1.11$
and  $\tau_{DIP} \simeq 1.33$ below and above $ {s}/ {s}_{\chi}=1$,
respectively. Notice also that the crossover is rather sharp, taking
place within one order of magnitude in $ {s}/ {s}_{\chi}$. This
is in contrast to the results of Ref. \cite{RYU-07}, where a large crossover
region with a slowly changing effective exponent was found, by using an
expression for the crossover avalanche size which is different from the
one found here. The crossover can also be seen by fitting a single power law
with an exponential cutoff,
\begin{equation}
\label{eq:single_PL}
P(s) = s^{-\tau_{eff}(  \lambda)}\exp\left(-\frac{ {s}}
{ {s}_0(  \lambda)}\right),
\end{equation}
to the data. The resulting effective exponent $\tau_{eff}$ as a function
of $  \lambda$ is shown in Fig. \ref{fig:crossover2} (b), showing again
a crossover between the values of $\tau_{LIM}=1.11$ and $\tau_{DIP}=1.33$.

To summarize, we have presented a theoretical analysis and a numerical model 
of DW morphology and avalanche dynamics in thin films with in-plane uniaxial 
anisotropy, giving rise to charged head-to-head (or tail-to-tail) DW's. As a 
result of the competition between DW surface tension and dipolar 
interactions, the DW's develop a zigzag structure. The avalanche dynamics 
displays a sharp crossover between two universality classes, characterized by 
the exponents $\tau_{LIM} \simeq 1.11$ and $\tau_{DIP} \simeq 1.33$, for scales 
dominated by the line tension and dipolar interactions, respectively. These two 
scaling regimes are separated by a crossover avalanche size $s_{\chi}$ which 
exhibits an exponential dependence on $\lambda \sim 1/M_s^2$. It is worth
noticing that the dipolar interactions scale as $ {q}^2 \log({q})$
in Fourier space. Hence, in the $q \rightarrow 0$ limit, the kernel is
similar to a negative surface tension, and it is 
therefore not possible to infer the dipolar universality class
based on simple power counting (as claimed e.g. in \cite{RYU-07}).
It would instead be necessary to perform a functional renormalization group 
calculation along the lines of Refs. \cite{NAT-92,NAR-93,LES-97,ERT-94,CHA-00,LED-02}, 
taking into account explicitly the non-convex nature of the interaction kernel,
leading to a violation of the no-passing rule usually obeyed by depinning
interfaces \cite{MID-92}.

{\bf Acknowledgments}. 
Claudio Serpico, Adil Mughal, James P. Sethna and Mikko Alava are thanked 
for interesting discussions. LL acknowledges the financial support of 
the Academy of Finland through a Postdoctoral Researcher's Project
(no. 139132) and via the Centres of Excellence Program (no. 251748),
as well as the computational resources provided by the Aalto Science-IT
project. SZ is supported by the European Research Council, 
AdG2001-SIZEFFECTS and thanks the visiting professor program of Aalto 
University.


\begin{thebibliography}{10}
\bibitem{DUR-06}
G. Durin and S. Zapperi, {\it The Barkhausen effect} in The Science of Hysteresis, 
edited by G. Bertotti and I. Mayergoyz, vol. II pp 181-267 (Academic Press, Amsterdam, 2006). 
\bibitem{SET-01}
J. P. Sethna, K. A. Dahmen, and C. R. Myers, Nature {\bf 410}, 242 (2001).
\bibitem{ZAP-98}
S. Zapperi, P. Cizeau, G. Durin, and H. E. Stanley, Phys. Rev. B {\bf 58}, 6563 (1998).
\bibitem{DUR-00}
G. Durin and S. Zapperi, Phys. Rev. Lett. {\bf 84}, 4705 (2000).
\bibitem{WIE-77}
N.~J. Wiegman,  Appl. Phys. \textbf{12}, 157 (1977).
\bibitem{WIE-78}
N.~J. Wiegman and R. Stege,  Appl.
Phys. \textbf{16}, 167 (1978).
\bibitem{PUP-00}
E.~Puppin, Phys. Rev. Lett., {\bf 84}, 5415 (2000).
\bibitem{KIM-03}
D.-H. Kim, S.-B. Choe, and S.-C. Shin, Phys. Rev. Lett., {\bf 90},  087203 (2003).
\bibitem{SAN-06}
L. Santi, F. Bohn, A.~D.~C. Viegas, G. Durin, A. Magni,R. Bonin,
S.  Zapperi and R. L. Sommer,
Physica B {\bf 384}, 144 (2006).
\bibitem{RYU-07}
K.-S. Ryu, H. Akinaga, and S.-C. Shin, Nature Physics {\bf 3}, 547 (2007).
\bibitem{SCH-04}
A. Schwarz, M. Liebmann, U. Kaiser, R. Wiesendanger, T. W. Noh, and D. W. Kim,
Phys. Rev. Lett. {\bf 92}, 077206 (2004).
\bibitem{LIE-05}
M. Liebmann, A. Schwarz, U. Kaiser, R.Wiesendanger, D.-W. Kim, and T.-W. Noh,
Phys. Rev. B {\bf 71}, 104431 (2005).
\bibitem{PAR-08}
S. S. P. Parkin, M. Hayashi, and L. Thomas, Science {\bf 320}, 190 (2008).
\bibitem{HAY-08}
M. Hayashi, L. Thomas, R. Moriya, C. Rettner, and S. P. Parkin, Science {\bf 320}, 209 (2008).
\bibitem{MUG-10}
A. Mughal, L. Laurson, G. Durin, and S. Zapperi, IEEE Trans. Magn. {\bf 46}, 228 (2010).
\bibitem{CER-06}
B. Cerruti and S. Zapperi, J. Stat. Mech. P08020 (2006).
\bibitem{CER-07}
B. Cerruti and S. Zapperi, Phys. Rev. B {\bf 75}, 064416 (2007).
\bibitem{BER-98}
G. Bertotti, {\it Hysteresis in Magnetism} (Academic Press, Boston, 1998).
\bibitem{RYU-06}
K.-S. Ryu, S.-C. Shin, H. Akinaga, and T. Manago, Appl. Phys. Lett. 
{\bf 88}, 122509 (2006).
\bibitem{TAL-11}
M. Talamali, V. Pet\"aj\"a, D. Vandembroucq, and S. Roux, Phys. Rev. E {\bf 84},
016115 (2011).
\bibitem{ROS-09}
A. Rosso, P. Le Doussal, and K. J. Wiese, Phys. Rev. B {\bf 80}, 144204 (2009).
\bibitem{NAT-92}
T. Nattermann, S. Stepanow, L.~H. Tang, and H. Leschhorn, J. Phys.
II (France)  {\bf 2},  1483  (1992).
\bibitem{NAR-93}
O. Narayan and D.~S. Fisher, Phys. Rev. B {\bf 48},  7030  (1993).
\bibitem{LES-97}
H. Leschhorn, T. Nattermann, S. Stepanow, and L.~H. Tang, Ann.
Physik {\bf 6},  1  (1997).
\bibitem{ERT-94}
D. Ertas and M. Kardar, Phys. Rev. E {\bf 49},  R2532  (1994).
\bibitem{CHA-00}
P. Chauve, T. Giamarchi, and P. Le Doussal
Phys. Rev. B 62, 6241-6267 (2000)
\bibitem{LED-02}
P. Le Doussal, K. J. Wiese, and P. Chauve
Phys. Rev. B 66, 174201 (2002)
\bibitem{MID-92}
A. A. Middleton, Phys. Rev. Lett. {\bf 68}, 670 (1992).
\end{thebibliography}
\end{document}